\documentclass[twocolumn,pra,superscriptaddress]{revtex4-1}
\usepackage{ natbib }
\usepackage{ amsmath,amssymb,}
\usepackage{graphicx}
\usepackage{dcolumn}
\usepackage{bm}

\usepackage[colorlinks,citecolor=red]{hyperref}

\begin{document}


\title{Simulating the Klein tunneling of pseudospin-one Maxwell particles with trapped ions}

\author{Peng He}
\affiliation{National Laboratory of Solid State Microstructures
and School of Physics, Nanjing University, Nanjing 210093, China}
\author{Xin Shen}%
\affiliation{
Shenzhen Institute for Quantum Science and Engineering
and Department of Physics, Southern University of Science and Technology, Shenzhen 518055, China
}%

\author{Dan-Wei Zhang}
\email{danweizhang@m.scnu.edu.cn}
\affiliation{Guangdong Provincial Key Laboratory of Quantum Engineering and Quantum Materials,
SPTE, South China Normal University, Guangzhou 510006, China}

\author{Shi-Liang Zhu}
	\email{slzhu@nju.edu.cn}
\affiliation{National Laboratory of Solid
State Microstructures and School of Physics, Nanjing University,
Nanjing 210093, China}

\affiliation{Guangdong Provincial Key Laboratory of Quantum
Engineering and Quantum Materials, SPTE, South China Normal
University, Guangzhou 510006, China}




\date{\today}

\begin{abstract}

We propose an experimental scheme to simulate and observe
the Klein tunneling of relativistic Maxwell particles with trapped ions. We explore the scattering dynamics of the pseudospin-one Maxwell particles and demonstrate that the scattered state should be a superposition of a reflection state, a localization state, and a transmission state. The probabilities of these states can be analytically obtained by
the approach of Landau-Zener transition. We further show that the Maxwell Hamiltonian and the associated scattering dynamics can be mimicked with two trapped ions. The Maxwell spinors are encoded by three internal states of the first ion, the position and momentum are described by those of the motional modes, and the desired linear potential barrier is built by the second ion.
\end{abstract}

\pacs{Valid PACS appear here}
\maketitle


\section{Introduction}

Simulation of the Dirac particles with condensed matter systems or some artificial systems has recently been
attracted considerable attention \cite{Wilczek1998,SLZhu2007,SLZhu2009,DWZhang2012b,Tarruell2012,Duca2015,DWZhang2012c,Gerritsma2010,Lamata2007,Casanova2010,Gerritsma2011,Casanova2011,SLiu2014}. The Dirac equation successfully merges quantum mechanics with special relativity, and predicts some unexpected peculiar effects for a relativistic quantum particle, such as Klein's paradox \cite{Klein1929,Calogeracos1999} and \emph{Zitterbewegung} \cite{Uber1930}. These predicted phenomena provide fundamental understanding of relativistic quantum effects, but are very hard to observe in elementary particles. In recent years, it has been demonstrated that various artificial quantum systems, such as trapped ions \cite{Gerritsma2010,Lamata2007,Casanova2010,Gerritsma2011} and ultracold atoms \cite{SLZhu2007,Tarruell2012,Duca2015,DWZhang2012b,DWZhang2012c}, can be used to simulate relativistic quantum effects. These systems become promising platforms for quantum simulation due to their high flexibility and controllability, which allow access to different physical regimes \cite{DWZhang2018,Leibfried2003,Hffner2008}.

Besides the spin-1/2 Dirac particles, exotic quantum effects can also appear in higher spin relativistic quantum particles \cite{LLiang2016,Urban2011,Betancur-Ocampo2017,XTan2018,AFang2016,HXu2017,ZHYang2016}. Similar to the spin-1/2 Dirac equation, a relativistic quantum wave equation can be formulated from the classical Maxwell equations. This wave equation describes the dynamics of an effective spin-1 relativistic quantum particle, and thus is called quantum Maxwell equation \cite{YQZhu2017,Oppenheimer1931}. It has been demonstrated that this quantum Maxwell equation can be simulated with ultracold atoms in optical lattices \cite{YQZhu2017,RShen2010,Dora2011,Goldman2011}. In its original formulation, the Klein tunneling was referred as an undamped scattering under a potential step with barrier hight $V > 2mc^2$, where $mc^2$ is the rest energy of the incident particle \cite{Klein1929,Calogeracos1999}. The Klein tunneling effect always comes with the transition and interference of different energy parts of states \cite{Thaller1992}.  Due to the richer energy structure for the spin-1 particles, perfect penetration against square barrier can be found for some special incident energy for the spin-1 particles, but less transmission for the spin-1/2 Dirac particles under the same conditions. Such phenomenon is termed as super-Klein tunneling in literature \cite{Urban2011,Betancur-Ocampo2017,AFang2016,HXu2017}. On the other hand, there
is one oscillation frequency in the \emph{Zitterbewegung} effect of the Dirac particles, but there are two different
oscillation frequencies in the \emph{Zitterbewegung} oscillations of Maxwell fermions \cite{XShen}. So both the Klein tunnelling and \emph{Zitterbewegung} effects for the quantum Maxwell particles can have unique features. However, the realization of the quantum Maxwell equation with cold atoms in optical lattices is challenge due to the complicated spin-orbit couplings required for the three-component spinors \cite{YQZhu2017}. Thus, other experimentally more feasible schemes to mimic the quantum Maxwell equation are highly desired.

In this paper, we propose an experimentally feasible scheme to simulate and observe the scattering dynamics described by the quantum Maxwell equation with trapped ions. We explore the scattering dynamics of the pseudospin-1 Maxwell particles in the presence of a linear external potential and demonstrate that the scattered state should be a superposition of a reflection state, a localization state, and a transmission state. The probabilities of these states can be analytically obtained by using
the approach of Landau-Zener transition. We further show that the Maxwell Hamiltonian and the associated scattering dynamics can be mimicked with two trapped ions, similar to the simulation of the Dirac Hamiltonian \cite{Gerritsma2010,Lamata2007,Casanova2010,Gerritsma2011}. The Maxwell spinors are encoded by the internal states of the first ion, and its position and momentum are described by those of the motional modes of the ion. The desired linear potential barrier is built by the second ion. Ions trapped in a RF trap can be well manipulated to deal with a wide range of information precessing tasks with high flexibility and accuracy \cite{Leibfried2003,Hffner2008}. The other experimental manipulations, such as preparation or readout, have already been the standard methods in trapped ion system. Notably, the Klein tunneling and \emph{Zitterbewegung} of a Dirac particle have been simulated and observed with trapped ions \cite{Gerritsma2011,Gerritsma2010}. The technologies developed there can be straightforwardly used in the present scheme. So the phenomena explored here could be observed within the near future.

The rest of the paper is organized as follows. In Sec. \ref{sec2}, we present the relativistic Hamiltonian of a pseudospin-1 Maxwell particle and reveal the probabilities of three scattering states in the Klein tunneling with the approach of Landau-Zener transition. In Sect. \ref{sec3}, we propose quantum simulation of the system and its scattering dynamics with trapped ions. Finally, a short conclusion is given in Sec. \ref{sec5}.

\section{The scattering dynamics}\label{sec2}

We consider quantum tunneling of pseudospin-1 relativistic particles through a potential $V(x,y)$ in a two-dimensional space, which is described the Hamiltonian
\begin{equation}
\hat H_M=c\hat p_xS_x+c\hat p_yS_y+mc^2S_z+V(x,y)I_3,\label{eq_model}
\end{equation}
where $c$ is the effective speed of light, $\hat p_{x,y}=i\hbar \partial_{x,y}$ are the momentum operators, and $m$ is the particle mass. The matrices $S_{x,y,z}$ are matrix representations of the spin components with spin $S=1$, and $I_3$ is the identity matrix. The Hamiltonian in Eq. (\ref{eq_model}) can be derived from the famous classical Maxwell equations in the forms of the Schrodinger equation \cite{YQZhu2017}, and it describes a (pseudo)spin-1 particle in relativistic case. Therefore, the particles described by the quantum wave equation $i\hbar\partial_t \Psi (\mathbf{r},t)=H_M\Psi(\mathbf{r},t)$ are so-called Maxwell particles \cite{YQZhu2017,Oppenheimer1931}.

A general state in momentum space $k_{x,y}=p_{x,y}/\hbar$ can be written as $|\Psi\rangle=\sum_{j=1}^3 a_{j}|\Psi_j\rangle$ with $j=+,-,0$, where $|\Psi_j\rangle$ are the eigenstates (spinors) with the eigenvalues $E_\pm=\pm\sqrt{c^2 k^2+m^2c^4}$ ($k=\sqrt{{k_x^2+k_y^2}}=p/\hbar$) and $E_0=0$. The spinor $|\Psi_j\rangle$  can be constructed by the projection operators in momentum space, $|\Psi_j\rangle=\hat P^j(k)|\Psi\rangle$, where the projection operators are given by
\begin{equation}
\begin{split}
\hat P^{\pm}=&\frac{1}{2}\big[(\frac{cp_xS_x+cp_yS_y+mc^2S_z}{E_+})^2\\ &~~~\pm\frac{cp_xS_x+cp_yS_y+mc^2S_z}{E_+}\big]\,,\\
\hat P^0=&I_3-(\frac{cp_xS_x+cp_yS_y+mc^2S_z}{E_+})^2.
\end{split}
\end{equation}

\begin{figure}[htbp]
\centering
\includegraphics[width=0.45\textwidth]{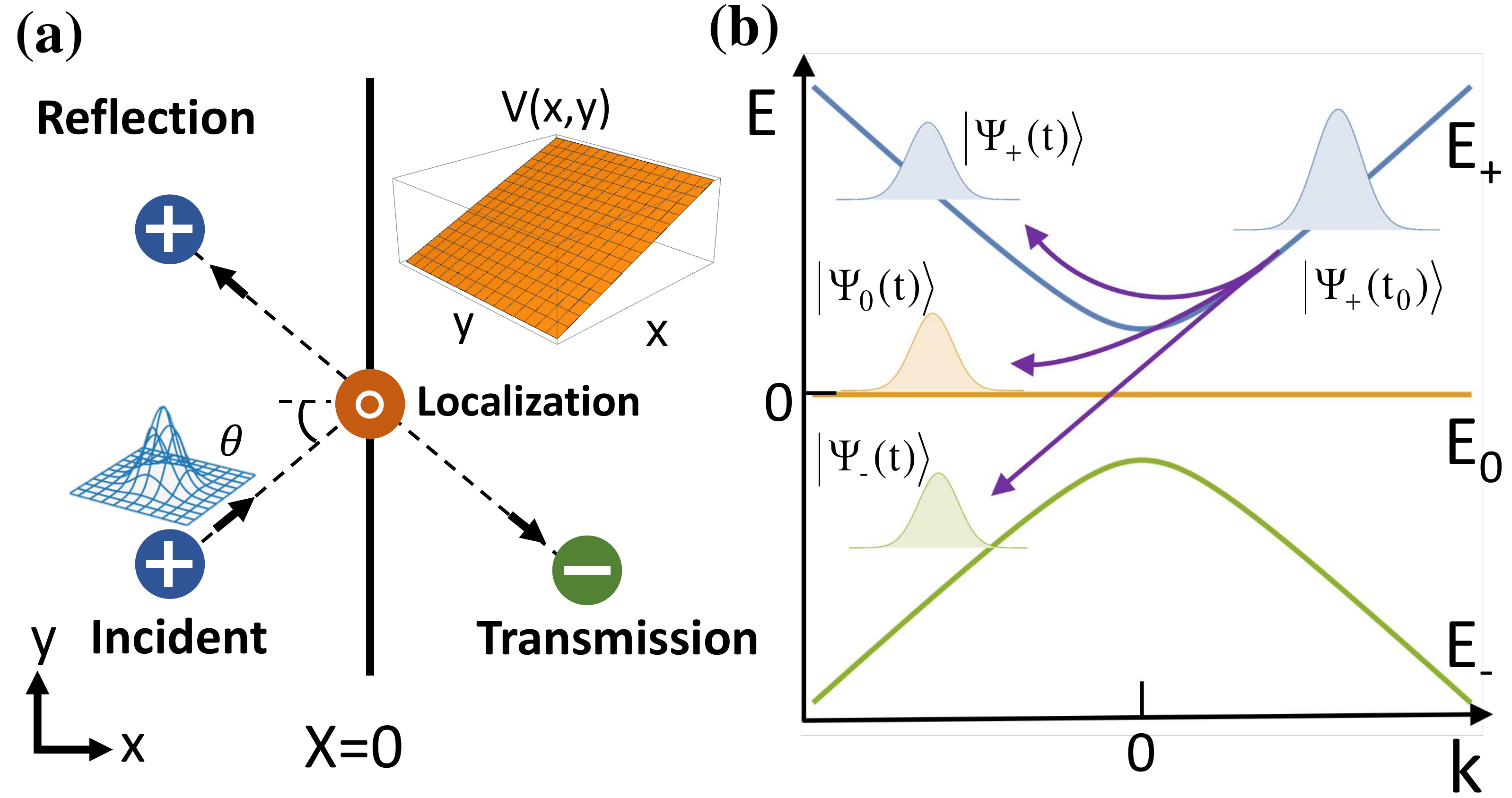}
\caption{Scattering dynamics of a pseudospin-1 Maxwell particle in (a) position space and (b) momentum space. The incident state $|\Psi_{+} (t_0)\rangle$ with positive momentum is initially prepared in the positive energy band. It moves in real space with a repulsive potential $V(x,y)=gx$ shown in the inset of (a). The particle can be reflected as a reflection wave, or enter the forbidden region as a transmission wave by reducing its kinetic energy, or even enter the zero-energy band as a localization state, as shown in (a). The three scattering states are denoted the wave functions $|\Psi_{+}(t)\rangle$, $|\Psi_{-}(t)\rangle$, and $|\Psi_{0}(t)\rangle$ in (b), respectively, and the distributions in the three bands can be calculated by the approach of Landau-Zener transition.}
\label{figone}
\end{figure}

Here we focus on the quantum tunneling with an external potential $V(x,y)=gx\ (g>0)$, as shown in Fig. \ref{figone}(a). If the initially incident state is $|\Psi_{j}\rangle$, the finial state after scattering can be written in a general form,
\begin{equation}
|\Psi(t)\rangle=\sum_{jn} a_{jn}|\Psi_n(t)\rangle,
\end{equation}
where $j,n=+,-,0$, and $a_{jn}$ represents the amplitude transfering from the initial state $|\Psi_{j}\rangle$ to the final energy branch of $|\Psi_{n}\rangle$.
In Fig. \ref{figone}(a), the case with the incident state $|\Psi_{+}\rangle$ is illustrated. The final state should be a superposition of a reflection state denoted as $|\Psi_{+}\rangle$, a localized state denoted as $|\Psi_{0}\rangle$, and a transmission state denoted as $|\Psi_{-}\rangle$. The state $|\Psi_{0}\rangle$ should be a localized state since the group velocities $\partial_{k_x,k_y}E_0 (k_x,k_y)=0$.

Under the linear potential condition, the transition probabilities $|a_{jn}^2|$ can be calculated based on the method of the Landau-Zener tunneling
\cite{Sauter1931,Landau1932,zener1932}. We rewrite the Hamiltonian (\ref{eq_model}) in momentum space,
\begin{equation} \label{eq_model_k}
\begin{split}
\hat H_k=&c\hbar k_x S_x+c\hbar k_yS_y+mc^2S_z+i\hbar g\partial_{k_x} I_3\\
=&c\hbar k_xS_x+\tilde mc^2\tilde S_{\tilde{z}}+i\hbar g\partial_{k_x}I_3,
\end{split}
\end{equation}
where $\tilde mc^2=\sqrt{m^2c^4+\hbar^2k_y^2c^2}$ is an effective mass since $k_y$ is well defined, and $\tilde S_{\tilde z}=n_yS_y+n_zS_z$ with $n_y=\hbar k_y/(\tilde mc)$ and $n_z=m/\tilde m$. The term $i\hbar g \partial_{k_x}$ is equivalent to a constant force along the $k_x$ axis, which makes a decrease in $k_x$. Thus, the scattering process can be interpreted in terms of a reduced Landau-Zener transition in one-dimensional momentum space (i.e., $k_x$). As shown in Fig. \ref{figone}(b), the incident state $|\Psi_{+} (t_0)\rangle$ has an initial momentum $\mathbf{k}_0=(k_{x0},k_{y0})$ with $k_{x0}>0$, and the linear potential decreases $k_x$ and leads to a non-adiabatic transition to another two bands near the anti-crossing point. The finite state acquires a reversed momentum along $k_x$, \emph{i.e.} $k_x <0$, which corresponds to a reflection in real space. Following the calculations outlined in Ref. \cite{Carroll1986}, we can obtain the transition probability

\begin{figure}[tbp]
	\centering
	\includegraphics[width=0.45\textwidth]{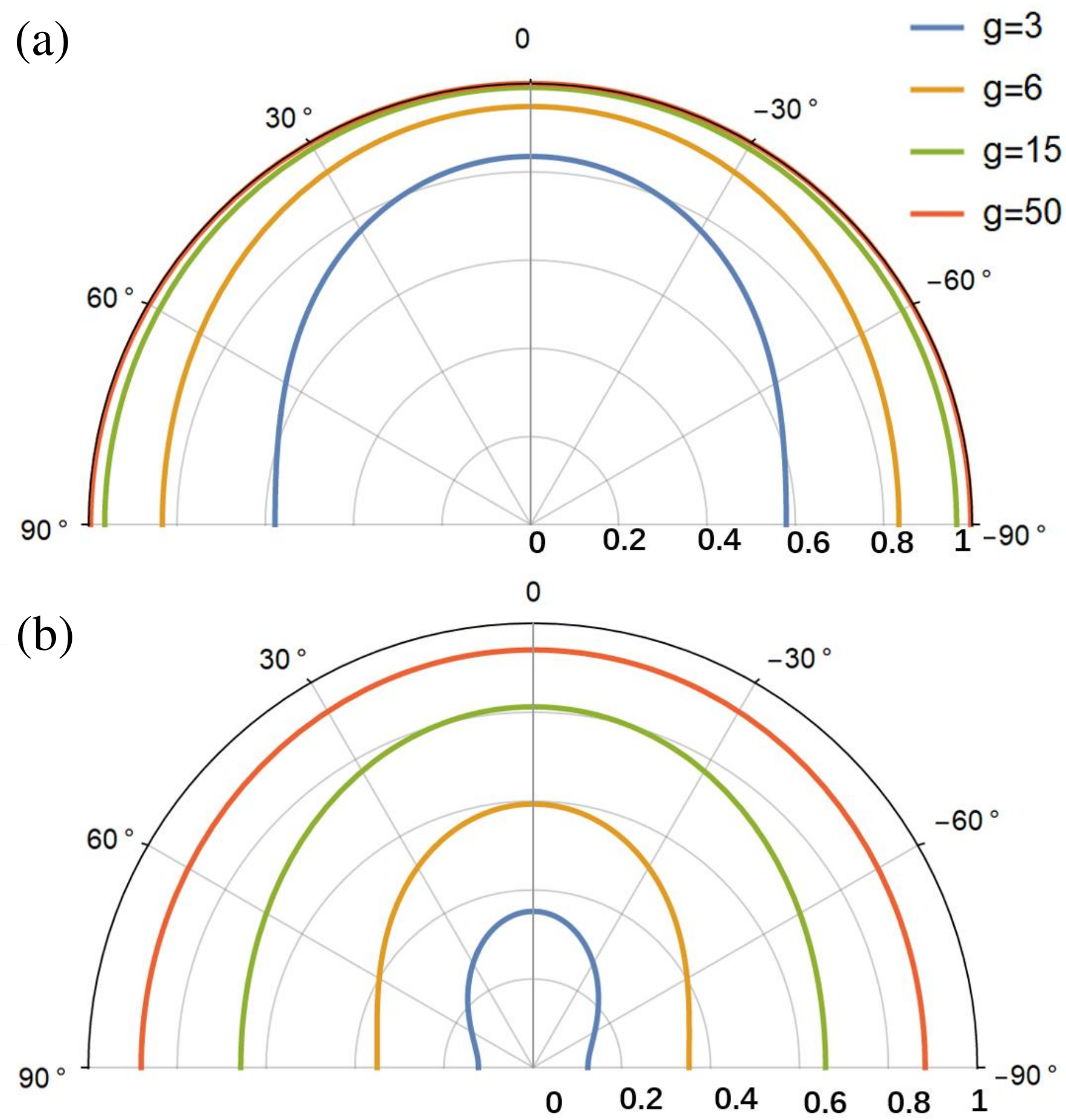}
	\caption{The transmission probability as a function of the incident angle $\theta=\arctan(k_{y0}/k_{x0})$ for (a) spin-1 particles and (b) spin-1/2 particles with different slope gradients $g$ of the linear potential. Here the natural units $\hbar=c=1$ are adopted, and the parameters $m=1$ and $k_{y0}=1$.}
	\label{figtwo}
\end{figure}

\begin{equation}
\begin{split}
\Gamma_{+-}=&|a_{+-}|^2=\exp(-\pi\frac{{\tilde m}^2c^4}{\hbar cg})\\=&\exp(-\pi\frac{m^2c^4+p_0^2c^2\sin^2\theta}{\hbar cg})\,,\\\Gamma_{+0}=&|a_{+0}|^2=2\exp(-\pi\frac{{\tilde m}^2c^4}{2\hbar cg})[1-\exp(-\pi\frac{\tilde{m}^2c^4}{2\hbar cg})]\\=&2\exp(-\pi\frac{m^2c^4+p_0^2c^2\sin^2\theta}{2\hbar cg})\times\\&[1-\exp(-\pi\frac{m^2c^4+p_0^2c^2\sin^2\theta}{2\hbar cg})]\,,\\
\Gamma_{++}=&|a_{++}|^2=1-\Gamma_{+-}-\Gamma_{+0}\,,\label{gamma}
\end{split}
\end{equation}
where $\Gamma_{jn}$ is the occupation probability on the $j$ band at time $t \rightarrow +\infty$ for the state initially on the $n$ band at $t \rightarrow -\infty$, and $\theta$ denotes the incident angle defined as $\theta=\arctan(k_{y0}/k_{x0})$. The transmission probability can be defined as
\begin{equation}
T=\Gamma_{+0}+\Gamma_{+-}.
\end{equation}

\begin{figure*}[htbp]
\centering
\includegraphics[width=0.9\textwidth]{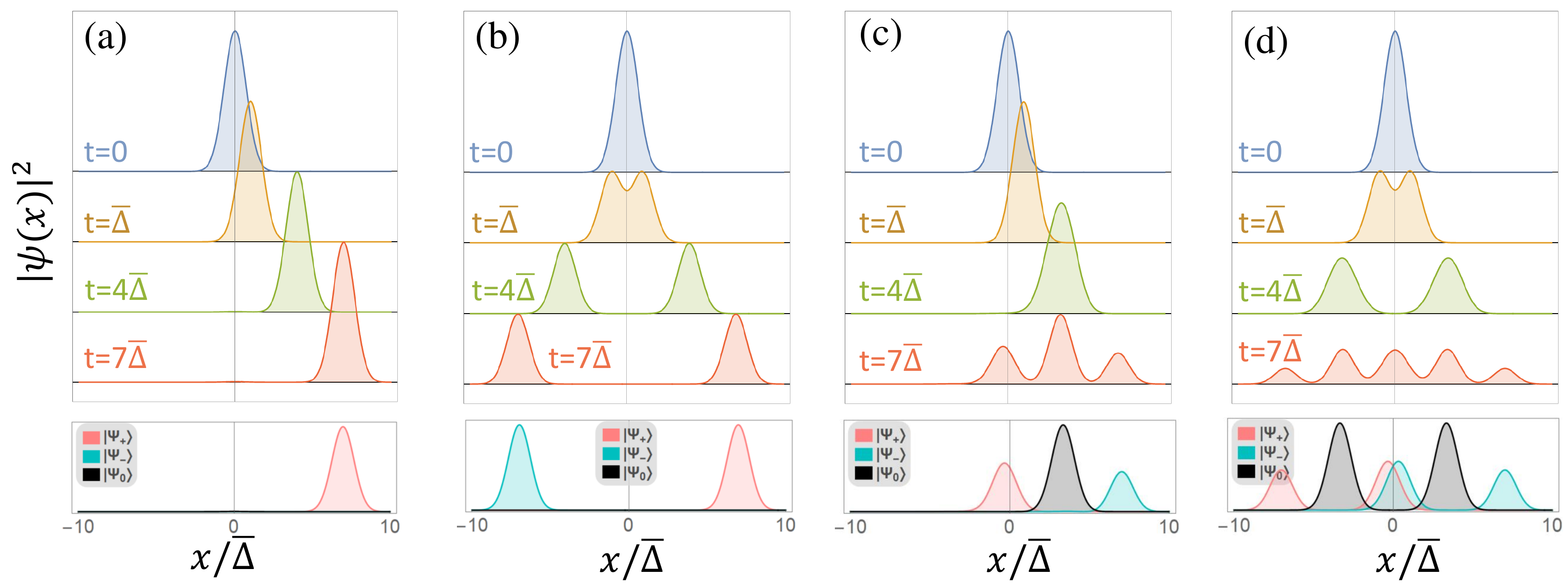}
\caption{Up panel: free evolution ($g=0$) for an initial wave packet with (a) a positive-energy eigen-spinor $|\Psi_+\rangle$ and (b) a superposition spinor $\frac{1}{\sqrt{2}}(|\Psi_+\rangle+|\Psi_-\rangle)$; the scattering dynamics with $g=1.5$ for an initial wave packet with (c) a positive-energy eigen-spinor $|\Psi_+\rangle$ and (d) a superposition spinor $\frac{1}{\sqrt{2}}(|\Psi_+\rangle+|\Psi_-\rangle)$. Down panel: the corresponding distributions of three spin components $|\Psi_{+,-,0} (t=7\bar\Delta)\rangle$ in real space. Here $\hbar=c=1$ and the rest mass $m=0.85$.}
	\label{figthree}
\end{figure*}

Figure \ref{figtwo}(a) illustrates the transmission probability $T$ as a function of the incident angle $\theta$ for the Maxwell particles. It shows that the potential will become more transparent for all the incident angles when the slope gradient $g$ becomes larger. If $g$ is sufficiently large, the transmission probability is almost unit and this result is very similar to the so-called super-Klein tunneling of spin-1 particles through a square potential barrier \cite{RShen2010,Urban2011}. For comparison, the transmission probability for spin-1/2 particles is also plotted in Fig. \ref{figtwo}(b). The relativistic Hamiltonian of spin-1/2 particles has the same form with that in Eq. (\ref{eq_model}), but the spin operators $S_{x,y,z}$ and the unit matrix $I_3$ are replaced by the Pauli matrices $\sigma_{x,y,z}$ and $2\times 2$ unit matrix $I_2$, respectively. It is clear that the transmission probability is smaller for spin-1/2 particles than that of spin-1 particles under the same conditions.

Without loss of generality, we consider the normal incident case for simulation of the scattering dynamics, in which case the model reduces to one dimension. Figures \ref{figthree}(a-d) show the scattering dynamics of the spin-1 particles described by an initial wave packet $\propto e^{ip_0x}e^{-\frac{x^2}{2\bar \Delta^2}}\xi$ with $p_0=10.0$ and $\bar\Delta=2$, where $\xi$ denotes the spinor function. Figures \ref{figthree}(a) and \ref{figthree}(c) plot the results for $\xi^{T}=(1,0,0)$ with $g=0$ (free evolution) and $g=1.5$ (scattering by the linear potential), respectively. Figures \ref{figthree}(b) and \ref{figthree}(d) show the results when the initial spinor state is $\xi^T=(1,0,1)$, with the same other conditions in \ref{figthree}(a) and \ref{figthree}(c), respectively. The down panel in Fig. \ref{figthree} show the corresponding distributions of three spin components $|\Psi_{j}(t=7\bar \Delta)\rangle $. For the case of $g=1.5$, the final state after a long time is a superposition of a reflection state $|\Psi_{+}\rangle$ propagating along the $-x$ direction (centering at the $x<0$ region), a localized state $|\Psi_{0}\rangle$ centering at the $x>0$ region, and a transmission state $|\Psi_{-}\rangle$ propagating along the $x$ direction (centering at the $x>0$ region), which are also shown in Fig. \ref{figone}(a). In Fig. \ref{figthree}(d), five peaks in the total density distribution are formed when the evolution time is longer than $7\bar \Delta$ because a pair of peaks appear for every $|\Psi_{j}\rangle$ ($j=+,-,0$) with two peaks of $|\Psi_{-}\rangle$ and $|\Psi_{+}\rangle$ almost being overlapped.

\section{Quantum simulation with trapped ions}\label{sec3}

\begin{figure}[tbp]
\centering
\includegraphics[width=0.35\textwidth]{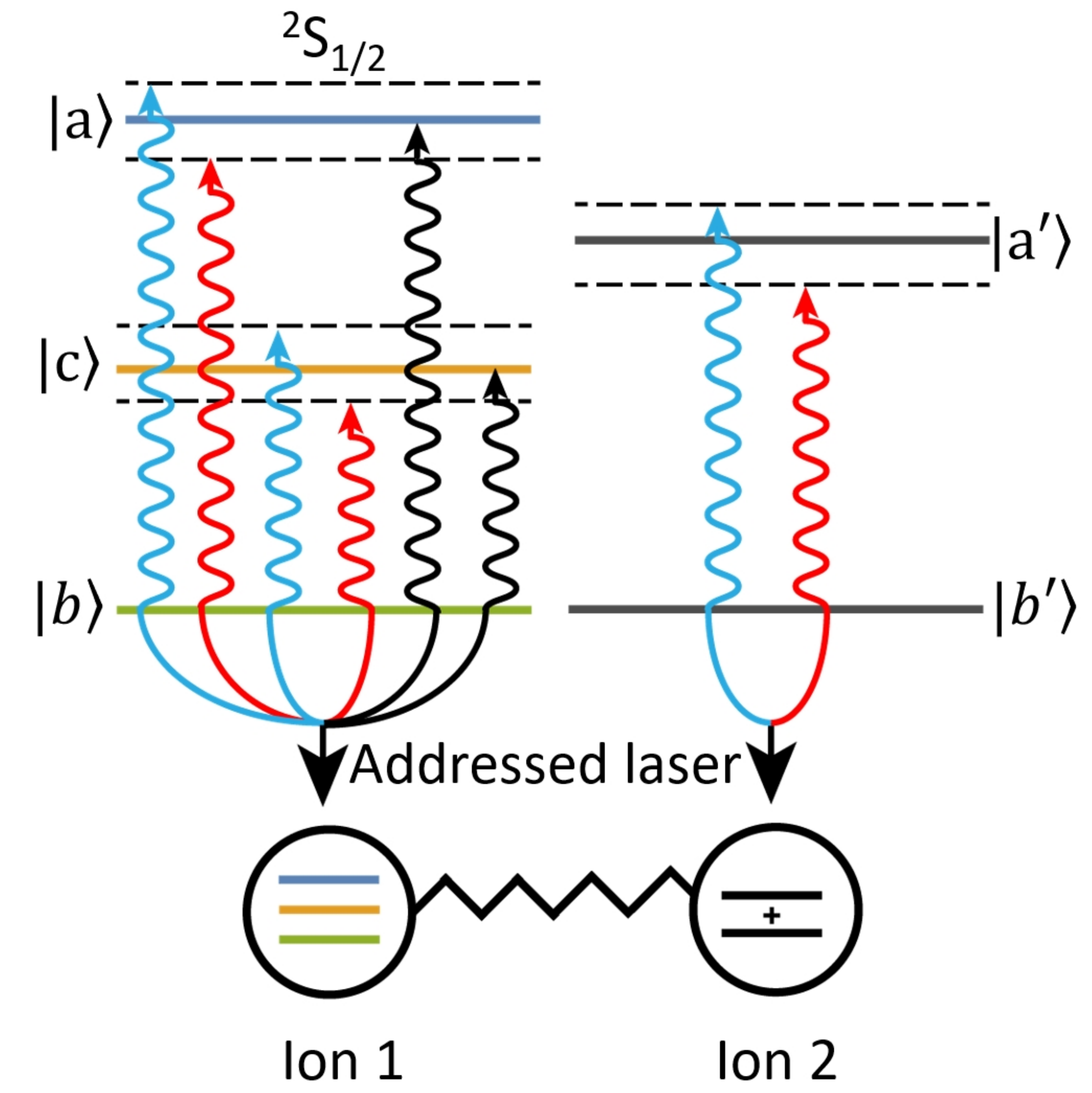}
\caption{Proposed trapped-ion setup for simulating scattering dynamics of a Maxwell particle in an external potential. The three-component spinors are encoded by three internal states of ion 1, and the potential is implemented by red and blue sidebands applied to the auxiliary ion 2.}
\label{setup}
\end{figure}

We now propose quantum simulation of scattering dynamics with trapped ions. Similar to the simulation of Dirac particles \cite{Casanova2010,Gerritsma2011}, mimicking the scattering process in a scale potential requires two ions. Let us consider a string of two trapped ions, denoted by ion 1 and ion 2, as shown in Fig. \ref{setup}. The first ion will encode a three-component Maxwell spinor in its three ionic states, while the second ion will be used as an ancilla to implement the potential. The involved internal levels for ion 1 could be chosen as the $^2S_{1/2}$ hyperfine clock ground states of $^{171}\rm{Yb}^+$:  $|a\rangle\equiv|1,1\rangle$, $|b\rangle\equiv|0,0\rangle$, and $|c\rangle\equiv |1,-1\rangle$; for ion 2 could be qubit states $|a'\rangle\equiv|1,0\rangle$ and $|b'\rangle\equiv|0,0\rangle$ \cite{Senko2015,Olmschenk2007}. Here the two quantum numbers denote $F$ and $m_F$ for the levels of $^{171}\rm{Yb}^+$ ion.

The motion of two ions trapped in a RF trap are correlated together to form collective modes. The creation and annihilation of motional modes give definition of external parameters by
\begin{equation}
\hat x_l=\Delta_l(\hat a_l+\hat a_l^\dagger)\,,\quad\hat p_l=\frac{\hbar}{2i\Delta_l}(\hat a_l-\hat a_l^\dagger)\,,\label{eq_ca}
\end{equation}
where $a_l$ ($a^\dagger_l$) is the creation (annihilation) operator of a motional mode in the $l$ axis with $l=\{x,y,z\}$, and $\Delta_l=\sqrt{\hbar/2M\omega_l}$ is the spread of ground state of the ion. Lasers or microwaves are applied to couple the internal levels of each ion and the center-of-mass motion in one of the three directions by absorbing or emission of photons  \cite{Leibfried2003,Hffner2008,SLZhu2006,SLZhu2006b,Kim2009}.

The dipole coupling between monochromatic light field and ions or Raman coupling between bichromatic light field and ions can be effectively described by a simple Rabi model within the Lamb-Dicke regime. In this case, the effective Hamiltonian of interactions between light and two qubit levels $|\alpha\rangle$ and $|\beta\rangle$ could take the lowest order in $\eta$ \cite{Leibfried2003,Hffner2008},
\begin{equation}
\hat H^{\alpha\beta}=\frac{\hbar}{2}\Omega_0\sigma^{+}_{\alpha\beta}\{1+i\eta(\hat ae^{-i\nu t}+\hat a^\dagger e^{i\nu t})\}e^{i(\phi-\delta t)}+\text{H.c.}\,,
\end{equation}
where $\sigma_{+}^{\alpha\beta}$ is the raising operator acting on the ion's internal states, $\Omega_0$ is the Rabi frequency, $\eta=k\sqrt{2M\omega}$ is the Lamb-Dicke parameter, $\nu$ is the frequency of the laser field, $\phi$ is the phase of the laser field, and $\delta$ is the laser detuning.

Depending on the detuning from the qubit transition, three types of fundamental interactions can be engineered as the quantum simulation toolbox to construct desired dynamics. The first is the case of resonance $\delta=0$ with the reduced Hamiltonian
\begin{equation}
\hat H_{c}^{\alpha\beta}=\frac{\hbar}{2} \Omega_0(\hat \sigma^{+}_{\alpha\beta}e^{i\phi}+\hat \sigma^{-}_{\alpha\beta}e^{-i\phi}).
\end{equation}
which is called carrier interaction. When the detuning is negative $\delta=-\nu$, the effective interaction is called red-sideband with
\begin{equation}
\hat H_{r}^{\alpha\beta}=\frac{\hbar}{2} \Omega_0\eta(\hat a\hat\sigma^{+}_{\alpha\beta}e^{i\phi}+\hat a^\dagger\hat \sigma^{-}_{\alpha\beta}e^{-i\phi}),
\end{equation}
while the blue-sideband for $\delta=+\nu$ with
\begin{equation}
\hat H_{b}^{\alpha\beta}=\frac{\hbar}{2} \Omega_0\eta(\hat a^\dagger\hat\sigma^{+}_{\alpha\beta}e^{i\phi}+\hat a\hat \sigma^{-}_{\alpha\beta}e^{-i\phi})\,.
\end{equation}
Interactions in the toolbox are adjustable with the phase $\phi$ and coupling strength $\Omega$. By recalling the definition of Eq. (\ref{eq_ca}) and making identifications that $\hat \sigma^x_{\alpha\beta}=(\hat \sigma^{+}_{\alpha\beta}+\hat \sigma^{-}_{\alpha\beta})/2$ and $\hat \sigma^y_{\alpha\beta}=(\hat \sigma^{+}_{\alpha\beta}-\hat\sigma^{-}_{\alpha\beta})/2i$, when only the carrier interaction is present, one has
\begin{equation}
\label{eq10}
\hat H_{\sigma_l}^{\alpha\beta}=\hbar \tilde \Omega_l \sigma_l^{\alpha\beta}
\end{equation}
with the appropriate phase setting. Applying the red-sideband interaction and blue-sideband interaction with appropriate manipulation of the laser filed, one could implement the coupling between the internal and external dimensions,
\begin{eqnarray}
\label{Hp}
\hat H_{p_l}^{\alpha\beta}&=&\Delta_l\Omega_{0l}\eta\hat \sigma_l^{\alpha\beta}\hat p_l,\\
 \hat H_{x_l}^{\alpha\beta}&=&\hbar \eta\Omega\sigma_{x_l}^{\alpha\beta}\hat{x} _l/\Delta_{x_l}. \label{Hx}
\end{eqnarray}

To simulate a spin-1 particle, three internal levels are needed to encode the three spinors. As shown in Fig. \ref{setup}, the three levels are denoted as $|a\rangle$, $|b\rangle$ and $|c\rangle$, and the spinor state can be expressed as
\begin{equation}
|\Psi\rangle\equiv \Psi_a|a\rangle+\Psi_b|b\rangle+\Psi_c|c\rangle=(\Psi_a,\Psi_b,\Psi_c)^{\rm{T}}\,.
\end{equation}
Applying Eqs. \eqref{eq10}-\eqref{Hx} appropriately to the three levels in ion 1 and the two levels in ion2, we can obtain the following Hamiltonian
\begin{equation}
\begin{split}
\hat H_{ion}=&\hat H_{p_x}^{ab}+\hat H_{p_x}^{bc}+\hat H_{p_y}^{ab}+\hat H_{p_y}^{bc}+\hat H_{\sigma_z}^{ab}+\hat H_{\sigma_z}^{bc}+\hat H_{x}^{a^\prime b^\prime},\\
=&\eta\Delta\tilde\Omega_1(\hat \sigma^{ab}_x+\hat \sigma_x^{bc})\hat p_x+\eta\Delta\tilde\Omega_1(\hat \sigma^{ab}_y+\hat \sigma_y^{bc})\hat p_y+\\&\hbar \Omega_1(\hat \sigma_z^{ab}+\hat \sigma_z^{bc})+\hbar \eta\tilde\Omega_2\hat x\sigma_2^x/\Delta\\
=&\sqrt{2}\eta\Delta\tilde{\Omega}_1 (S_x\hat p_x+S_y \hat p_y)+\hbar \Omega_1 S_z+\hbar \eta\tilde\Omega_2\hat x\sigma_2^x/\Delta\,,\label{eq_eh}
\end{split}
\end{equation}
where $\hat H_{x}^{a^\prime b^\prime}$ and the related $\sigma_2^x$ correspond to the manipulations on ion-2. Here we have chosen the parameters so as $\eta:=\eta_{1x,y}=\eta_2$, $\tilde\Omega_1:=\Omega_{1x}=\Omega_{1y}$, $\Omega_1=\Omega_{1z}$ and $\Delta:=\Delta_{1}=\Delta_{2}$. Note that the $\sigma_z$ terms could be induced by AC Stark shift. To obtain a scalar potential which is linear in the $x$-direction, the second ion is required to be initialized in the positive eigenstate of $\sigma_2^x$. Then we can find that it is equivalent to the model given by Eq. (\ref{eq_model}) with the correspondence $c:=\sqrt{2}\eta\Delta\tilde\Omega_1$, $mc^2:=\hbar \Omega_1$ and $g:=\hbar\eta \tilde \Omega_2/\Delta$. Thus the two-dimensional quantum Maxwell equation can be realized with tunable parameters.

In absence of interactions $\hat H_{p_y}^{ab}$ and $\hat H_{p_y}^{bc}$, the Hamiltonian Eq. \eqref{eq_eh} reduces to one dimension, which could be more easily dealt with in realistic experiments. Under this condition, the scattering dynamics shown in Fig. \ref{figthree} can be experimentally demonstrated.

In typical experiments, the $^{171}\rm{Yb}^+$ ions can be initialized to the ground state by Doppler cooling on the $^2S_{1/2}$-$^2P_{1/2}$ transition at $369.53~\rm nm$. Then the carrier interaction and red/blue sideband interactions can be applied simultaneously to produce the desired dynamics via two-photon stimulated Raman transitions. Afterward, the ion states are measured by the standard flourescence technique with an additional laser field coupling the $^2S_{1/2}$-$^2P_{1/2}$ transition \cite{Olmschenk2007,Kim2009}. Note that the accessible parameters in experiments is sufficient to give observable phenomena. For instance, with typical experimental parameters given by $\eta=0.05$, $\tilde \Omega_1=2\pi \times 10~\rm{kHz}$, $\tilde \Omega_1=2\pi \times 1~\rm{kHz}$ and $\tilde \Omega_2=2\pi \times 50~\rm{kHz}$, which correspond to $m^2c^4/(\hbar c g) \sim 0.56$ in Hamiltonian \ref{eq_model_k}, one could prepare a normal incident positive-energy initial state with spatial extension of order of $\Delta$ to obtain a large transmission probability $T\approx 0.65$. Such tunneling process is expected to take place within about 1 ms, which is well within the typical decoherence time.

\section{Conclusions}\label{sec5}

In summary, we have proposed a feasible scheme to simulate and detect the scattering dynamics of the quantum Maxwell equation with trapped ions. We have shown that the Klein tunnelling probabilities of the Maxwell particles in the linear potential can be resolved by the Landau-Zener transition in momentum space. We have demonstrated that the Klein tunneling can be observed in trapped ion experiments based on quantum simulation of the Maxwell particles with the ion modes. Notably, simulation of the Klein tunneling in the Dirac equation has been experimentally achieved with trapped ions \cite{Gerritsma2011,Gerritsma2010}, and the technologies developed there can be directly used in our proposed scheme. Thus, the present scheme is quite promising for realizing the first experiment on simulation of the Maxwell particles.

\acknowledgments
 This work was supported by the NKRDP of China (Grant No. 2016YFA0301803), the NSFC (Grants  No.91636218,  No. 11604103, No. U1801661, No. and 11474153), the NSAF (Grant No. U1830111), the KPST of Guangzhou (Grant No. 201804020055), the NSF of Guangdong Province (Grant No. 2016A030313436), and the Startup Foundation of SCNU.


\end{document}